\documentclass[aps,prl,twocolumn,superscriptaddress,showpacs,groupedaddress]{revtex4-1}

\usepackage{amsmath,amssymb,graphicx,color,textcomp}

\newcommand{\bra}[1]{\langle#1|}
\newcommand{\ket}[1]{|#1\rangle}

\begin{document}

\title{Optically erasing disorder in semiconductor microcavities with dynamic nuclear polarization}

\author{T. C. H. Liew}
\author{V. Savona}
\affiliation{Institute of Theoretical Physics, Ecole Polytechnique F\'{e}d\'{e}rale de Lausanne EPFL, CH-1015 Lausanne, Switzerland}

\begin{abstract}
The mean squared value of the photonic disorder is found to be reduced by a factor of $100$ in a typical GaAs based microcavity, when exposed to a circularly polarized continuous wave optical pump {\it without} any special spatial patterning. Resonant excitation of the cavity mode excites a spatially non-uniform distribution of spin-polarized electrons, which depends on the photonic disorder profile. Electrons transfer spin to nuclei via the hyperfine contact interaction, inducing a long-living Overhauser magnetic field able to modify the potential of exciton-polaritons.
\end{abstract}

\date{\today}


\maketitle

Fermi's contact hyperfine interaction in solid-state systems allows an electron's spin polarization to be transferred to a nucleus~\cite{Overhauser1953,Lampel1968}. For this reason the optical excitation of spin polarized electrons can induce a dynamic nuclear polarization in bulk semiconductors~\cite{Lampel1968} as well as in quantum well (QW)~\cite{Flinn1990,Barrett1994,Marohn1995} and quantum dot~\cite{Brown1996,Gammon2001,Braun2006} structures. This mechanism allows for enhanced nuclear magnetic resonance imaging of nanostructures~\cite{Flinn1990,Barrett1994,Marohn1995,Gammon2001}, strengthened electron coherence for quantum spintronics~\cite{Brown1996,Burkard1999,Reilly2008}, and transfer of the electron spin to nuclear spin states for a long lived quantum memory~\cite{Imamoglu2003,Taylor2003}.

In a so far rather separate line of research, semiconductor microcavities~\cite{Kavokin2007} strongly couple the electronic states of QWs to light creating new eigenstates known as exciton-polaritons, which are perhaps most famous for their Bose-Einstein condensation~\cite{Kasprzak2006}. These systems are inevitably affected by disorder, arising from strain due to the lattice mismatch between layers~\cite{Gurioli2000,Langbein2002,Savona2007}. Disorder results in a fine structure in the energy of polariton condensates~\cite{Krizhanovskii2009} and optical parametric oscillators~\cite{Krizhanovskii2006}. Additionally, disorder inhibits the observation of the Berezinski-Kosterlitz-Thouless phase transition; although vortices have been observed~\cite{Lagoudakis2008} they can be attributed to scattering with disorder~\cite{Liew2008a}.

Recently, there has been a trend towards the control of the polariton potential~\cite{IdrissiKaitouni2006,Balili2007,Lai2007,Bajoni2007,Wertz2010} experienced by polaritons. In particular it has been shown that it is possible to engineer the polariton landscape with an optical excitation~\cite{Amo2010a}. In principle, through a very specific spatial patterning of the optical excitation this can counteract the disorder. The suppression of disorder would lower the threshold for polariton condensation~\cite{Kasprzak2006,Balili2007,Lai2007} and polariton lasing~\cite{Christopoulos2007}, and also enhance the characteristics of information processing devices based on ballistic polariton propagation~\cite{Johne2010,Laussy2010} or polariton neurons~\cite{Liew2008b,Amo2010b}. Unfortunately, disorder potentials naturally have an intricate, and mostly unknown, structure making the necessary patterning of the optical field a challenging task.

\begin{figure}[h]
\centering
\includegraphics[width=8.116cm]{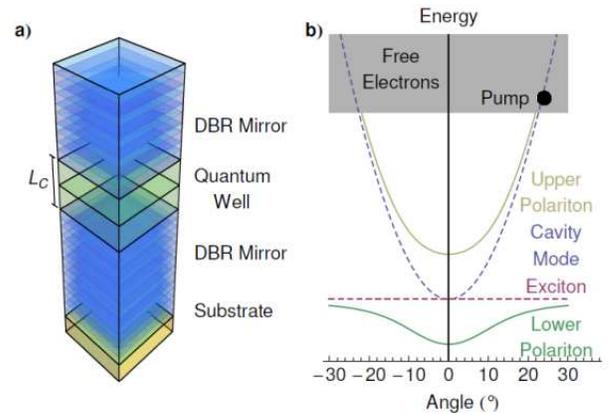}
\caption{a) Schematic of a semiconductor microcavity containing a QW. b) Dispersion of the different modes of the system. Excitons are strongly coupled to a cavity mode forming the polariton branches. Our scheme considers an optical pump exciting free electron-hole pairs at an energy near resonant with the cavity photon branch.}
\label{fig:scheme}
\end{figure}

Here, we show that one can screen the polariton disorder potential with an optical field that does not require any special spatial patterning. There are three steps involved in our method: 1) We consider the injection of spin polarized QW electrons with an optical excitation tuned just above the QW bandgap energy ($\sim1.52eV$ for a GaAs quantum well) and at an angle resonant with the cavity photon branch (see Fig.~\ref{fig:scheme}). Note that electrons are not excited in other layers of the structure, which have a larger bandgap (typically, the cavity has a bandgap $\sim1.58eV$ and the Bragg mirror layers are larger still~\cite{Langbein2002,Amo2010a,Amo2010b}). The cavity photon experiences a similar disorder profile to the LPB, and with careful choice of the laser energy, one can arrange for more electrons to be excited where the disorder potential is minimum and less where it is maximum; we naturally obtain a spatial pattern in the electron spin distribution. The electrons are free, yet typically do not move significant distances within their lifetime compared to the scale of photonic disorder, which is of the order of a few $\mu m$~\cite{Savona2007}.

2) The electrons can transfer their spin polarization to nuclei via the hyperfine interaction (note that there is no significant coupling of hole spins to nuclei since the valence wavefunctions are p-like~\cite{Flinn1990}). We consider the case of undoped samples and high excitation densities. As nuclear spins orient, a magnetic field develops in the system - the Overhauser field~\cite{Overhauser1953}. The Zeeman splitting of a nuclear spin is $3$ orders of magnitude smaller than that of an electron, such that an excess energy is generated when the electron spin is transferred to a nucleus. In quantum dot systems this limits the first-order electron-nuclear spin flip interaction, requiring second order processes in which a phonon carries away the excess energy~\cite{Erlingsson2001}. However, in QWs it has been noted that the continuous spectrum of electron states allows the excess energy to be carried away by the electron~\cite{Erlingsson2001}. We calculate the scattering rate using Fermi's golden rule and show that this process can result in a significant nuclear spin polarization at sufficiently high electron densities. 3) The enhanced Overhauser field shifts the potential of polaritons and, if it has a suitable distribution, compensates and screens the photonic disorder. At this stage the optical pump can be removed and the nuclear spin profile survives with a very slow diffusion rate ($0.1\mu m$ in $0.75$s)~\cite{Kalevich2008} and long lifetime (at least of the order of ms~\cite{Braun2006,Maletinsky2007} and possibly up to minutes~\cite{Krapf1990}). These timescales are very long compared to the ps-scale polariton lifetime~\cite{Amo2010b}, which determines the timescale of typical microcavity experiments. In the rest of this Letter, we consider the three steps described above in detail.

The hyperfine interaction of a single electron acting on a single nuclear spin has the Hamiltonian~\cite{Gammon2001,Erlingsson2001,Imamoglu2003}:
\begin{equation}
\mathcal{H}_\mathrm{hf}=\nu_0 A \left|\psi(\mathbf{R})\right|^2\left(\hat{I}_x\hat{S}_x+\hat{I}_y\hat{S}_y+\hat{I}_z\hat{S}_z\right),\label{eq:Ham}
\end{equation}
where $\nu_0$ is the unit cell volume and $A$ is the hyperfine coupling constant (averaged over Ga and As isotopes). This parameter depends on the nuclear and electronic properties of the material and is well-known experimentally~\cite{Gammon2001,Braun2006,Imamoglu2003}. $\psi(\mathbf{R})$ represents the electron envelope function, evaluated at the position of the nuclear spin. The lattice periodic Bloch function is absorbed in the definition of $A$. $\mathbf{\hat{I}}$ and $\mathbf{\hat{S}}$ represent the nuclear and electron spin operators. Considering spins quantized in the $z$ direction, we note that only the first two terms are responsible for spin flips; the operators $\hat{I}_z$ and $\hat{S}_z$ do not change the $z$ component of nuclear or electron spins.

Fermi's golden rule gives the nuclear spin-flip rate:
\begin{equation}
\Gamma_\mathrm{hf}=\frac{2\pi}{\hbar}\sum_{f,i}\left|\bra{\Psi_f}\mathcal{H}_\mathrm{hf}\ket{\Psi_i}\right|^2,\label{eq:GamHF}
\end{equation}
where the sum is over initial and final states, $\ket{\Psi_i}$ and $\ket{\Psi_f}$ respectively. The states $\ket{\Psi_{i,f}}=\ket{S_{i,f}}\ket{I_{i,f}}$ can be separated into the electron spin wavefunction, $\ket{S_{i,f}}$, and the nuclear spin wavefunction, $\ket{I_{i,f}}$. Taking the electron envelope function, appearing in Eq.~\ref{eq:Ham}, as a plane wave in the QW plane and the fundamental mode in the growth direction, the overlap integral in the QW center is:
\begin{equation}
\bra{\Psi_f}\mathcal{H}_\mathrm{hf}\ket{\Psi_i}=\frac{2\nu_0 A}{L V_{2D}}\bra{S_f,I_f}\hat{I}_x\hat{S}_x+\hat{I}_y\hat{S}_y\ket{S_i,I_i}.
\end{equation}
$L$ is the QW width and $V_{2D}$ is a normalization area.

The electron spin-relaxation time due to non-nuclear processes ($8$ns~\cite{Flinn1990}) is typically much longer than the electron lifetime ($\sim250$ ps~\cite{Flinn1990}); electrons can recombine radiatively with holes either directly or through the formation of excitons as intermediate states. For this reason we assume that the majority of electrons are completely polarized. Note that the hyperfine interaction is too weak to significantly affect the average electron spin polarization, given the much slower rate. Moreover, in the steady-state, recombination and spin relaxation of electrons is
compensated by fresh spin-polarized electrons, replenished by the pump. The sum over initial states in Eq.~\ref{eq:GamHF} is replaced by the number of electrons in the normalization area and a sum over nuclear spin states; $\sum_i=n_eV_{2D}\sum_{I_i}$, where $n_e$ is the 2D electron density. The sum over final states is made by an integration over electron states with in-plane wavevector $\mathbf{k}_\parallel$:
\begin{align}
\Gamma_\mathrm{hf}&=\frac{2\pi}{\hbar}\left(\frac{2\nu_0 A}{L V_{2D}}\right)^2n_eV_{2D}\frac{I(I+1)}{3}\notag\\
&\hspace{20mm}\times\frac{V_{2D}}{\left(2\pi\right)^2}\int d\mathbf{k}_\parallel\delta\left(E_i-E_f\right).
\end{align}
$E_i$ and $E_f$ represent the initial and final energies. We used the result of Ref.~\cite{Erlingsson2001} that $\sum_{I_f,I_i}|\bra{S^\prime}\bra{I^\prime}\hat{I}_x\hat{S}_x+\hat{I}_y\hat{S}_y\ket{S}\ket{I}|^2=I(I+1)/3$, where $I$ denotes the total nuclear spin. $^{69}$Ga, $^{71}$Ga and $^{75}$As all carry $I=3/2$.

Taking a parabolic dispersion for electrons, with effective mass $m_e$, we find:
\begin{equation}
\Gamma_\mathrm{hf}=\frac{n_em_e}{\hbar^3}\left(\frac{2\nu_0 A}{L}\right)^2\frac{I(I+1)}{3}.
\end{equation}
We note that to balance the loss in Zeeman energy, electrons have scattered to higher momentum states. However, since the 2D density of electron states in energy is constant, the rate is independent of the amount of energy transferred. For a typical GaAs microcavity, the spin-flip rate is very slow compared to the electron lifetime (for $n_e=10^{12}$cm$^{-2}$, $\Gamma_\mathrm{hf}\sim50$s$^{-1}$); as mentioned before the contact hyperfine interaction has negligible effect on the electron spin distribution. Nevertheless, when compared to the long nuclear spin relaxation time, high electron densities can result in a significant nuclear polarization.

We now consider optical excitation of the cavity branch by a spatially uniform circularly polarized continuous wave pump laser (Fig.~\ref{fig:scheme}b). The electron density depends on the cavity photon-pump detuning, which varies along the microcavity plane due to the disorder:
\begin{equation}
n_e(\mathbf{x}_\parallel)=\frac{F\Gamma^2}{\left(E_C(\mathbf{x}_\parallel)-E_p\right)^2+\Gamma^2}.
\end{equation}
$F$ represents the pump intensity, $\Gamma$ is the linewidth (HWHM), $E_p$ is the laser energy and $E_C(\mathbf{x_\parallel})$ is the energy of the cavity mode at the pump wavevector.

The average nuclear spin polarization is given by the rate equation~\cite{Kalevich2008,Braun2006}:
\begin{equation}
\frac{d\langle I_z (\mathbf{x}_\parallel)\rangle}{dt}=\Gamma_\mathrm{hf}(\mathbf{x}_\parallel)\left(Q\langle S_z\rangle-\langle I_z(\mathbf{x}_\parallel)\rangle\right)-\frac{\langle I_z (\mathbf{x}_\parallel)\rangle}{T_d}\label{eq:Izdiff}
\end{equation}
where $T_d$ is the nuclear spin relaxation time. The quantity $Q=\frac{I(I+1)}{S(S+1)}$, where $S=1/2$ represents the total spin of electrons. We have neglected the possible diffusion of nuclear spins, which can be avoided given the long timescale for nuclear spin diffusion~\cite{Kalevich2008}. The nuclear spin diffusion should be further suppressed by the Knight field of the strongly polarized electrons in our system~\cite{Lai2006}.

\begin{figure}[h]
\centering
\includegraphics[width=8.116cm]{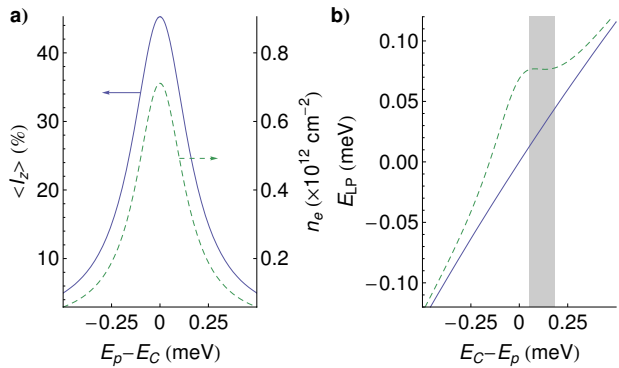}
\caption{a) Dependence of the steady state nuclear polarization (blue solid curve), as a fraction of complete polarization, on the detuning between the cavity photon energy and pump energy. The dashed green curve shows the excited electron density. b) The dependence of the lower polariton energy without (blue solid curve) and with (green dashed curve) the influence of dynamically polarized nuclei on the cavity photon-pump detuning. In the range marked in grey $E_{LP}(\mathbf{x}_\parallel)$ is roughly constant with respect to $E_C(\mathbf{x}_\parallel)$. Parameters: $L=15$nm, $\nu_0=(5.65\AA)^3/2$, $A=90\mu$eV~\cite{Imamoglu2003} (this represents an average for $^{69}$Ga, $^{71}$Ga and $^{75}$As nuclei), $Q=5$~\cite{Kalevich2008,Braun2006}, $T_d=10$ms, $\Gamma=0.12$meV, $V=2.55$meV~\cite{Amo2010b}. The (average) photon-exciton detuning was taken as $3$meV.}
\label{fig:GammaHF}
\end{figure}

Numerical solution of Eq.~\ref{eq:Izdiff} shows that steady states are reached within $\sim10$ms. The steady state solution is $\langle I_z(\mathbf{x}_\parallel)\rangle=\frac{Q\langle S_z\rangle\Gamma_\mathrm{hf}(\mathbf{x}_\parallel)T_d}{1+\Gamma_\mathrm{hf}(\mathbf{x}_\parallel)T_d}$. The dependence of the steady state values of $\langle I_z(\mathbf{x}_\parallel)\rangle$ on the local pump detuning $E_p-E_C(\mathbf{x}_\parallel)$ is shown in Fig.~\ref{fig:GammaHF}a, together with the excited electron density. The considered densities are within the achievable limits of semiconductor microcavities~\cite{Laussy2010}.

The energies of lower branch polaritons, $E_{LP}(\mathbf{x}_\parallel)$, are given by the standard two-mode coupling equation:
\begin{align}
E_{LP}(\mathbf{x}_\parallel)&=\frac{_1}{^2}\left(E_{C^\prime}(\mathbf{x}_\parallel)+E_X(\mathbf{x}_\parallel)\right.\notag\\
&\hspace{10mm}\left.-\sqrt{\left(E_{C^\prime}(\mathbf{x}_\parallel)-E_X(\mathbf{x}_\parallel)\right)^2+4V^2}\right)
\end{align}
where $E_{C^\prime}$ is the energy of the cavity mode at zero in-plane wavevector, $E_X$ is the exciton energy, and $V$ is the exciton-photon coupling constant. The exciton energy is influenced by the Overhauser magnetic field according to $E_X(\mathbf{x}_\parallel)=E_{X^\prime}+A \langle I_z(\mathbf{x}_\parallel) \rangle$, where $E_{X^\prime}$ is the exciton energy in the absence of any nuclear polarization (note that the shift is given by the same hyperfine coupling constant that enters the Hamiltonian in Eq.~\ref{eq:Ham}, which already contains the electron g-factor~\cite{Coish2009}). A typical dependence of the polariton energy on the cavity photon-pump detuning is shown in Fig.~\ref{fig:GammaHF}b. The modification due to the nuclear polarization results in a part of the curve where $E_{LP}(\mathbf{x}_\parallel)$ is almost constant with varying $E_C(\mathbf{x}_\parallel)$, as marked by the grey rectangle. We can therefore expect that if the photonic disorder varies within the region marked by the grey rectangle, which can be tuned by varying $E_p$, then $E_{LP}(\mathbf{x}_\parallel)$ will experience a reduced disorder potential.
\begin{figure}[h]
\centering
\includegraphics[width=8.116cm]{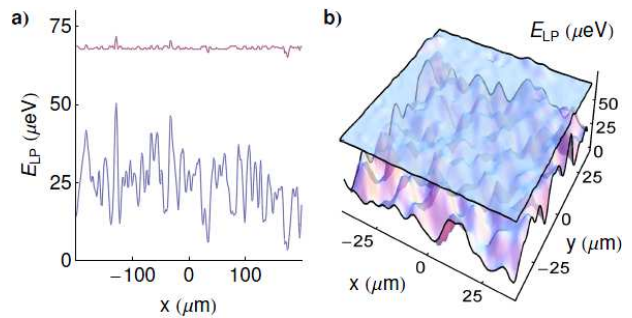}
\caption{Typical variation of the lower polariton energy in space without (lower curve in (a) and lower surface in (b)) and with (upper curve and upper surface) the influence of nuclear spins. The photonic disorder was modelled with a correlation length of $2\mu$m and rms amplitude $0.04$meV.}
\label{fig:DisorderScreening}
\end{figure}

We model a typical microcavity with a Gauss correlated disorder, described by a mean squared amplitude and correlation length~\cite{Savona2007} (although we represent disorder in this way, our conclusions hold for any shape of the disorder profile). Figure~\ref{fig:DisorderScreening} shows the effect that the nuclear spins can have on the disorder potential experienced by polaritons. In our calculations the mean squared amplitude of the disorder was reduced by a factor of $100$.

The results of this paper are derived considering realistic parameters for GaAs based microcavities. Larger Overhauser energy shifts could be obtained in structures containing Indium, which has a total spin $I=9/2$. Another possibility would be to make use of the polariton-magnetic ion coupling in dilute magnetic semiconductors (such as CdMnTe cavities). However, it is notable that magnetic ions typically exhibit a faster spin relaxation.

We point out that in our scheme it is the true potential experienced by polaritons that is modified rather than the introduction of an effective potential due to polariton-polariton interactions. This is different to the work of Ref.~\cite{Amo2010a} in which one distribution of polaritons modifies the potential of another. In that case, one can still have scattering between the polaritons creating the potential and those being influenced. Whilst our scheme is only capable of screening the disorder for one spin polarization, there are no secondary effects on polaritons. Both heat and electronic excitations leave the system on a much faster time scale than the nuclear spin relaxation, once the optical pump is turned off. Heat has been calculated to propagate through GaAs at a rate of $40 \mu m$ in $712ps$ in Ref.~\cite{Cotta2007}. The typical width of a microcavity (in the growth direction) is smaller, about $5\mu m$~\cite{Kavokin2007}, and we have verified by solving the heat equation with the parameters of Ref.~\cite{Cotta2007} that heat can be completely dissipated from a sample within a few $100$s of $ps$. Quasi-continuous wave excitation would also be feasible for reducing heating~\cite{Balili2007}.

With regard to the binding of electrons and holes into excitons, the rate is given by the bimolecular formation coefficient $C\approx10cm^2/s$~\cite{Piermarochi1997}. Solving the rate equation $dn_e/dt=-Cn_e^2$, shows that an initial electron density of $n_e=10^{12} cm^{-2}$ drops by a factor of $10,000$ within $1 ns$. Excitons decay radiatively in about $10ps$ in GaAs~\cite{Kavokin2007}.

In summary, whilst disorder in semiconductor microcavities has stimulated interesting studies~\cite{Langbein2002,Savona2007} and was exploited in spin current separation~\cite{Leyder2007}, disorder is usually a nuisance in experiments. Disorder is inevitable, raises the threshold for polariton condensation and restricts polariton ballistic transport. We present a scheme to compensate that disorder. The excitation of a typical microcavity with an unpatterned optical field induces a nuclear spin polarization. The resulting Overhauser magnetic field screens the polariton disorder potential by a factor of $100$ in a typical GaAs based microcavity. The nuclear spin distribution has a lifetime far exceeding that of typical microcavity experiments and so can remain after the removal of the optical pump, allowing plenty of time for further experiments, manipulation or application of exciton-polaritons in a disorder-free environment.

We thank M. Wouters, W. Langbein \& I. Carusotto for encouraging discussions. Our work was supported by NCCR Quantum Photonics (NCCR QP), research instrument of the Swiss National Science Foundation (SNSF).

\end{document}